\documentclass[english, floatfix, prl, preprint, groupedaddress, superscriptaddress]{revtex4-1}
\usepackage{babel}
\usepackage{verbatim}
\usepackage{bm} 
\usepackage{amsmath}
\usepackage{graphicx}
\usepackage{esint}
\usepackage[T1]{fontenc}
\usepackage[latin9]{inputenc}
\usepackage{color}

\makeatletter
\usepackage{physics}
\usepackage{bbm}
\usepackage{bbold}

\newcommand{\Z}{\mathcal{Z}}%
\newcommand{\T}{\mathcal{T}}%
\newcommand{\A}{\mathcal{A}}%
\newcommand{\B}{\mathcal{B}}%

\makeatother

\begin{document}
\title{Pre-Cooling Strategy Allows Exponentially Faster Heating}
\author{A. Gal}
\author{O. Raz}
\affiliation{Department of Physics of Complex Systems, Weizmann Institute of Science, 76100 Rehovot, Israel}

\begin{abstract}
    What is the fastest way to heat a system which is coupled to a temperature controlled oven? The intuitive answer is to use only the hottest temperature available. However, we show that often it is possible to achieve an exponentially faster heating, and propose a strategy to find the optimal protocol. Surprisingly, this protocol can have a pre-cooling stage -- cooling the system before heating it shortens the heating time significantly. This approach can be applied to many-body systems, as we demonstrate in the 2D antiferromagnet Ising model.
\end{abstract}

\maketitle


Consider the common task of cooling a hot system by coupling it to a thermal reservoir with a controlled temperature, as a refrigerator. It is counter-intuitive but well understood that a preceding heating stage followed by a slow cooling stage often shorten the overall cooling time. Indeed, \emph{annealing} techniques are widely used in industrial treatment of metals, glasses and crystal lattices \citep{dossett2006practical,humphreys2012recrystallization}, where the pre-heating stage accelerates the relaxation to equilibrium by decreasing the number of dislocations in the material and relieving internal stresses. A similar approach is used in {simulated annealing} \citep{kirkpatrick1983optimization,ingber1996adaptive,de2003placement,xu2018entropy}. These \emph{Monte Carlo} (MC) optimization algorithms find an approximation of the global minimum of a function, using an artificial temperature which characterizes the probability to accept a step to a state with a different value of this function. In order to escape local minima, the temperature is initially set to a high value, then slowly decreased.
Another non-monotonic relaxation phenomenon is the \emph{Mpemba effect} (ME) \cite{mpemba1969cool,AmJPhys_General_jeng2006mpemba,MpembaWaterLasanta,Granular_Mpemba_PhysRevLett,LargeMpemba,QuantumMpemba,SpinGlassMpemba}, where an initially hot system cools faster than an identical system initiated at a lower temperature. In contrast to annealing, where the heating can be fast but the cooling must be slow, to observe a ME the temperature of the bath must be lowered instantaneously. Although the ME seems to suggest that a pre-heating stage can shorten cooling processes, it is not necessarily the case since the preceding stage might take a longer time than gained.

Are there cases, in analogy to the cooling optimization problem, where it is faster to heat a system by first cooling it? Improving the heating rate by changing other variables was already concerned in the \emph{shortcut to adiabaticity} literature \cite{martinez2016engineered,tu2014stochastic,abah2018performance} and is relevant in many applications. For example, shortening the heating stroke period in a heat engine can improve its power output \cite{brandner2015thermodynamics,tu2014stochastic,abah2018performance,Pugatch_PRL_engines}. In analogy to the ME, the recently introduced \emph{inverse Mpemba effect} (IME) \cite{lu2017nonequilibrium}, where a cold system heats faster than an identical system initiated at a warmer temperature when both are quenched to a high temperature bath, suggests that this might be possible. Nevertheless, it does not necessarily imply that pre-cooling speeds up heating for a similar reasoning as in the ME -- the cooling stage might take a longer time than gained by the IME.

In this manuscript we show that a pre-cooling strategy can result in an exponentially faster heating. After formulating the problem, we propose a strategy to construct optimal heating protocols and demonstrate it in a specific diffusion problem. In this example a pre-cooling protocol speeds up heating exponentially in a system that does not exhibit any variant of the IME, demonstrating that such protocols are not necessarily a consequence of the IME, and are expected in a wider range of systems. 
To address many-body systems and avoid intractable calculations, we then extend our strategy by a projection of the dynamics into a lower dimension space. This approach is demonstrated in the 2D antiferromagnet Ising model.



\begin{figure}
    \centering
    \includegraphics[width=1\columnwidth]{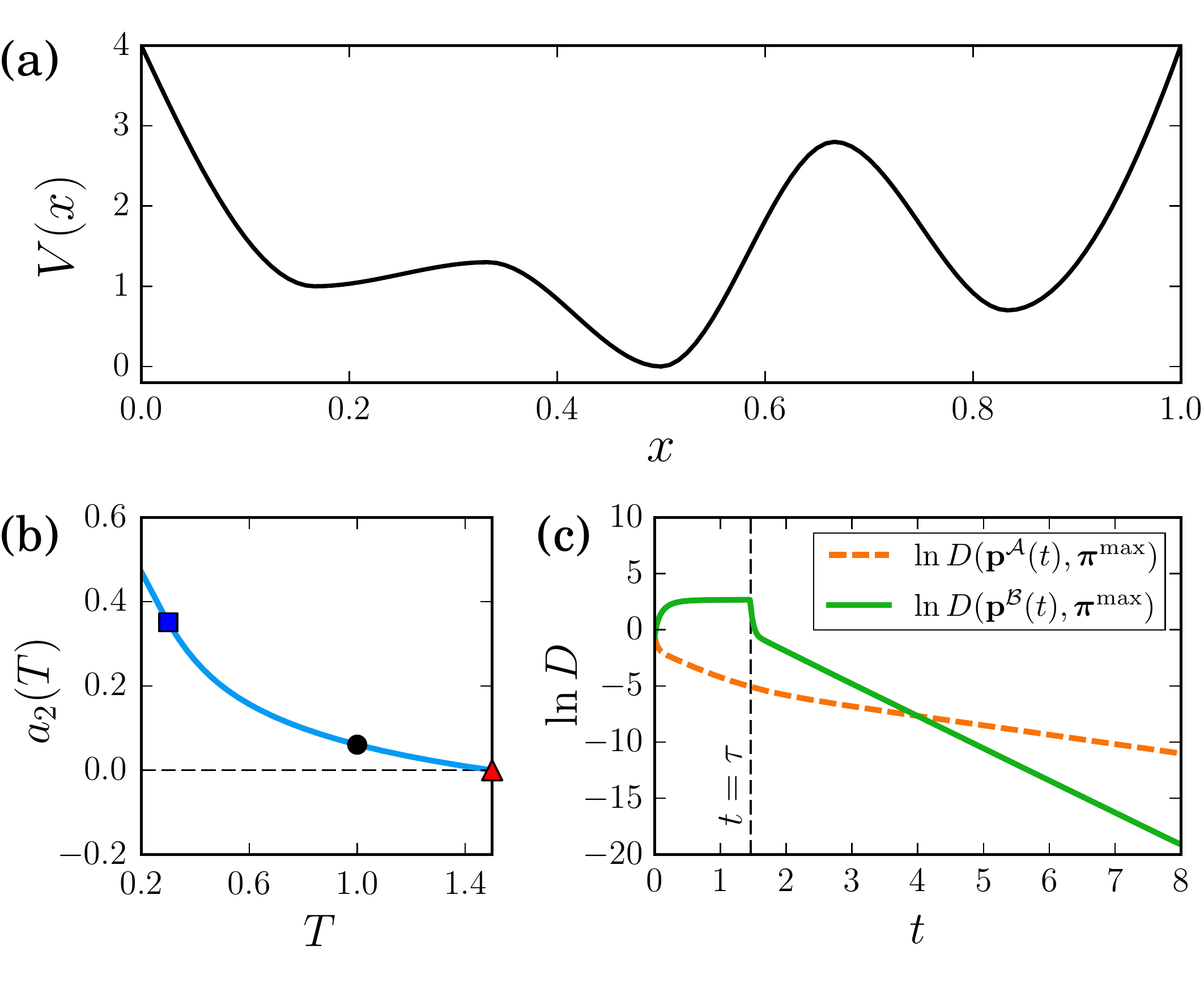}
    \caption{A Brownian particle in a potential.
    (a)
    The potential $V(x)$.
    (b)
    The coefficient $a_2(T)$ of the slowest direction ${\bf v}_2$ in the equilibrium distribution $\pi(x;T) = \boldsymbol{\pi}(T)$. The red square is at $T_\textrm{max}=1.5$ where $a_2(T_\textrm{max}) = 0$, the black point is at the initial temperature $T_0=1$ and the blue triangle is at the cold temperature $T_{\textrm{cold}}=0.3$. The coefficient $a_2(T)$ is monotonic, therefore the system does not exhibit any type of an IME.
    (c)
    The log-distance (KL divergence) to the equilibrium $\boldsymbol{\pi}(T_\textrm{max}) = \boldsymbol{\pi}^\textrm{max}$ of the trajectories generated by the oven protocol ($\A$, dashed orange) and the pre-cooling protocol ($\B$, solid green), both initiated at $\boldsymbol{\pi}(T_0)$. The pre-cooling duration is $\tau=1.46$, after which the pre-cooling protocol achieves an exponentially faster relaxation rate towards equilibrium.}
    \label{fig:FokkerPlanckExample}
\end{figure}

To define ``shorter heating time'', we next introduce the mathematical setup. For simplicity let us first consider systems with $N$ states. Let $p_i(t)$ denotes the probability to find the system in state $i$ at time $t$. A probability distribution of an $N$-state system is represented by ${\bf p}=(p_1,\ldots,p_N)$, with $0\leq p_i\leq 1$ and $\sum_{i=1}^{N}p_i=1$. The thermal bath is assumed to have zero memory, and thus the dynamic of ${\bf p}(t)$ is Markovian and follows the master equation
\begin{equation}\label{eq:master_equation}
    \dot{\bf p}(t) = R(T_b){\bf p}(t).
\end{equation}

\noindent The transition rate from state $j$ to state $i$ is given by the matrix element $R_{ij}(T_b)$ where $T_b$ is the bath temperature. The negative diagonal elements $R_{ii}(T_b) = -\sum_{j\neq i} R_{ji}(T_b)$ are the escape rates from state $i$. As $R(T_b)$ describes relaxation towards equilibrium, it is detailed balanced and its equilibrium probability distribution, denoted by ${\boldsymbol\pi}(T_b)$, is given by the Boltzmann distribution: 
\begin{equation}\label{eq:bolzmann_distribution}
    \pi_i(T_b) = \frac{1}{\Z}e^{-E_i/T_b},\;\;\; \Z = \sum_i e^{-E_i/T_b},
\end{equation}
where $E_i$ is the energy of state $i$ and $T_b$ is in units where $k_B=1$. By writing $R(T_b)$ we assume that the only degree of control at our disposal is the bath temperature. The explicit functional dependence of $R_{ij}$ on $T_b$ does not play a significant role in our analysis, as long as its equilibrium is given by Eq.~(\ref{eq:bolzmann_distribution}).

We consider heating processes in which the system is initiated at the equilibrium ${\boldsymbol\pi}(T_0)$ for a specific temperature $T_0<T_\textrm{max}$, where $T_\textrm{max}$ is the maximal temperature of the bath. Our goal is to heat the system towards the hot equilibrium ${\boldsymbol\pi}(T_\textrm{max})$. The dynamic is defined by the heating protocol $T_b(t)$ -- bath temperature as a function of time -- which we limit by $T_b(t)\leq T_\textrm{max}$. The trajectory in the probability space, generated by $T_b(t)$, is given by 
\begin{equation}\label{Eq:TimeOrder}
    {\bf p}(t) = \T\left\{e^{\intop_0^t R\left(T_b(t')\right)dt'}\right\}{\boldsymbol\pi}(T_0),
\end{equation}
where $\T$ is the time-ordering operator.

In what follows we compare two types of heating protocols. The first is the \emph{oven} protocol, where the system is heated by a time independent temperature, $T_b(t) = T_\textrm{max}$. The second is an \emph{alternative} heating protocol, that during the time interval $t\in[0,\tau]$ is constrained by $T_b(t)\leq T_\textrm{max}$, and for $t>\tau$ we assume that the bath temperature is set to $T_\textrm{max}$. 

To gain some insight on the heating process, it is beneficial to decompose ${\bf p}(t)$ in terms of the right eigenvectors of $R(T_\textrm{max})$. Let ${\bf v}_i$ be a solution of 
\begin{equation}
    R(T_\textrm{max}){\bf v}_i = \lambda_i {\bf v}_i,
\end{equation} 
where $0=\lambda_1>\lambda_2\geq \lambda_3\geq...\geq\lambda_N$ are the (sorted) eigenvalues of $R(T_\textrm{max})$, which are real valued as $R(T_\textrm{max})$ is detailed balanced \cite{gaveau1997general}. The trajectories ${\bf p}(t)$ in probability space can be expressed as:
\begin{align}\label{eq:trajectory_components}
    {\bf p}(t) = {\boldsymbol\pi}(T_\textrm{max})+\sum_{i=2}^N a_i(t){\bf v}_i.
\end{align}
For $t>\tau$, the rate matrix is fixed since $T(t) =T_\textrm{max}$ in both of the protocol types, and the dynamic is simplified to
\begin{equation}\label{Eq:CompDecay}
    a_i(t) = a_i(\tau)e^{\lambda_i (t-\tau)},
\end{equation}
where $a_i(\tau)$ are determined by the protocol $T_b(t)$. For the oven protocol, the dynamic is even simpler,
\begin{equation}\label{eq:ComponentDecayOven}
    a_i(t) = a_i^0e^{\lambda_i t},
\end{equation}
where $a_i^0$ is the coefficient of ${\bf v}_i$ in $\boldsymbol{\pi}(T_0)$. 

A naive approach to optimize heating protocols is to choose some distance function $D({\bf p}(\tau), {\boldsymbol\pi}(T_\textrm{max}))$ that measures the distance of ${\bf p}(\tau)$ to ${\boldsymbol\pi}(T_\textrm{max})$, e.g. the Kullback-Leibler (KL) divergence \citep{lu2017nonequilibrium,jarzynski2011equalities,still2012thermodynamics}, and find $T_b(t)$ that minimizes this distance \footnote{We assume that this distance does not vanish for any finite time protocol, due to the constraint $T(t)\leq T_\textrm{max}$.}. However, a key point in our analysis is that the relaxation does not stop at $t=\tau$. The coupling with the $T_\textrm{max}$ bath continues to drive the system towards ${\boldsymbol\pi}(T_\textrm{max})$ for $t>\tau$, hence the distance to equilibrium at $t=\tau$ is not a good objective to minimize (See, e.g., Fig.~\ref{fig:FokkerPlanckExample}c).

The strategy we suggest, following Eqs.~(\ref{eq:trajectory_components},\ref{Eq:CompDecay}), is that instead of minimizing the distance to equilibrium at $t=\tau$ we should minimize the magnitude of $a_i(\tau)$ by their order. The exponential time dependence in Eq.~(\ref{Eq:CompDecay}) implies that at long enough time the dominant component in Eq.~(\ref{eq:trajectory_components}) is the slowest one. Consequently, $a_2(t)$ dominates the distance to equilibrium relaxation, regardless of the values of $a_i(\tau)$ for $i>2$. Therefore, the optimal protocol is the one that minimizes $a_2(\tau)$. If there are several protocols for which $a_2(\tau)=0$, then among these we should choose the protocol that minimizes $a_3(\tau)$, and so on. In other words, the optimal protocol sets $0=a_2(\tau)=a_3(\tau)=...=a_m(\tau)$ for the largest $m$ possible and minimizes $a_{m+1}(\tau)$, leading to the fastest relaxation towards equilibrium.

The above strategy can be readily generalized to any detail balanced Markovian system with a discrete set of eigenvalues, and as we show in the Supplemental Material \cite{SI}, even for non-linear dynamics. Let us demonstrate our approach by considering the following example: a Brownian particle diffusing in a 1D potential with reflecting boundary conditions, described by the Fokker-Planck equation
\begin{equation} \label{eq:FokkerPlanck}
     \dot{p}(x,t) = \mathcal{L}(T) p(x,t),
\end{equation}
where $p(x,t) \equiv {\bf p}(t)$ is the probability distribution of finding the particle in position $x\in (0,1)$ at a given time $t$, and
\begin{equation}
    \mathcal{L}(T)p(x,t) \equiv \frac{1}{\gamma}\partial_x\Big{[}\big{(}\partial_x V(x)\big{)}p(x,t)\Big{]} + \frac{T}{\gamma}\partial_x^2 p(x,t).
\end{equation}

Finding the optimal protocol $T_b(t)$ for this system is very challenging. However, a pre-cooling stage with a single cold temperature is enough to set $a_2=0$, and therefore to exponentially improve the relaxation time.  Such a pre-cooling protocol is illustrated in Fig.~(\ref{fig:FokkerPlanckExample}). For simplicity we assume that the damping coefficient is given by $\gamma = 1$. Specifically, $V(x)$ is chosen to be the potential plotted in Fig.~\ref{fig:FokkerPlanckExample}(a). We next compare two protocols, both initiated at the equilibrium $\boldsymbol{\pi}(T_0)$, but evolving under different $T(t)$: (i) In the oven protocol (orange dashed line),  the temperature is set to $T_\textrm{max}$ at all times; (ii) In the pre-cooling protocol (green solid line), the system is first coupled to a $T_\textrm{cold}$ bath for a finite duration $\tau$, and then to the $T_\textrm{max}$ bath. Fig.~\ref{fig:FokkerPlanckExample}(c) shows the KL divergence of $p(x,t)$ to the equilibrium $\boldsymbol{\pi}(T_\textrm{max})$ as a function of time. 
During the pre-cooling stage the distance to $\boldsymbol{\pi}(T_\textrm{max})$ increases while $a_2(t)$ decreases and vanishes at $t=\tau$. For $t>\tau$, where the system is coupled to the $T_\textrm{max}$ bath, $a_2(t) = 0$ and the slowest dynamics in the system towards $\boldsymbol{\pi}(T_\textrm{max})$ does not take part in the relaxation process. Therefore, it relaxes exponentially faster towards its equilibrium, as can be seen from the different slopes of the log-distance in Fig.~\ref{fig:FokkerPlanckExample}(c). As shown in Fig.~\ref{fig:FokkerPlanckExample}(b), for this potential $a_2(T)$ is monotonic with temperature, so the system does not show any type of an IME. Nevertheless, heating can be improved by pre-cooling.

Under what conditions can pre-cooling improve heating? As demonstrated above, this happens even in systems that do not show any type of IME. However, in the limited case of systems that exhibit a \emph{strong inverse Mpemba effect} (SIME), a simple argument for the existence of such a protocol can be given. The SIME is defined by the existence of a temperature $T_M < T_\textrm{max}$ at which $a_2(T)$ changes its sign \cite{vucelja2017mpemba}. If this effect exists in the system, then the oven protocol is necessarily not optimal for any initial temperature $T_M < T_0 < T_\textrm{max}$ where $a_2(T_0)\ne0$. Pre-cooling the system to temperature $T_\textrm{cold}<T_M$ initiates a trajectory from ${\boldsymbol\pi}(T_0)$ towards ${\boldsymbol\pi}(T_\textrm{cold})$, where both of these two equilibrium points have a different sign of $a_2$,
therefore $a_2(t)$ must cross the $a_2=0$ manifold in a finite time. This time is chosen as $\tau$ for the pre-cooling protocol. In other words, the SIME assures that a pre-cooling protocol can be constructed to eliminate $a_2$ at a finite time and hence improve the heating rate. An example that provides a physical intuition for this limited case is given in the Supplemental Material \cite{SI}.


In the analysis above, the Markovian operator and its second eigenvector ${\bf v}_2$ played a crucial role. It is therefore rarely applicable to many-body systems, where the number of microstates grows exponentially with the number of particles and thus finding ${\bf v}_2$ or ${\boldsymbol\pi}(T_0)$ is a highly nontrivial task. To overcome this limitation, we next extend our strategy by considering a projection of the high-dimensional probability space trajectories into a lower dimension space.

Given a many-body system, we first choose two different observables, $x_1$ and $x_2$, that can be easily calculated for any microstate of the system. A probability distribution ${\bf p}$ can then be projected into a 2D space by the ${\bf p}$-averaging of $x_1$ and $x_2$ over all microstates, given by $(\langle x_1\rangle_{\bf p},\langle x_2\rangle_{\bf p})$. Whereas it is impractical to follow the time evolution of ${\bf p}(t)$ in a system with a huge number of microstates, $\langle x_1\rangle_{{\bf p}(t)}$ and $\langle x_2\rangle_{{\bf p}(t)}$ can be evaluated to a high precision using a standard MC simulation. As discussed above, in the full probability space all trajectories ${\bf p}(t)$ asymptotically approach $\boldsymbol{\pi}(T_\textrm{max})$ from the slowest direction ${\bf v}_2$, except for ones that are on the fast manifold. 
Therefore, their projections $(\langle x_1\rangle_{{\bf p}(t)},\langle x_2\rangle_{{\bf p}(t)})$ approach the mapped equilibrium $(\langle x_1\rangle_{\boldsymbol{\pi}(T_\textrm{max})},\langle x_2\rangle_{\boldsymbol{\pi}(T_\textrm{max})})$ from the projection of ${\bf v}_2$ direction \footnote{We assume that the projection of ${\bf v}_2$ into the 2D space is non-zero. If it is zero, then a different set of observables $x_1$ and $x_2$ can be chosen.}. In contrast, trajectories on the fast manifold approach $\boldsymbol{\pi}(T_\textrm{max})$ from a different direction ${\bf v}_i$ $(i\ge3)$, and projected to trajectories that approach the mapped equilibrium from the projection of the ${\bf v}_i$ direction \footnote{In the rare cases where the projections of ${\bf v}_2$ and ${\bf v}_i$ happens to coincide, a different set of observables $x_1$ and $x_2$ must be chosen.}.

To demonstrate the projection in a concrete example of a many-body system, we consider the 2D Ising model on an $N\times N$ square lattice, with external magnetic field, antiferromagnetic nearest neighbor interactions and periodic boundary conditions \footnote{Antiferromagnetic interactions and periodic boundary conditions are consistent only for even $N$.}. We denote the state of the spin located at the $i^{th}$ row and $j^{th}$ column by $\sigma_{ij}\in\{-1,1\}$. The Hamiltonian of the system is given by 
\begin{equation}
    \mathcal{H} = - \frac{J}{4}\sum_{\langle ij,kl \rangle}\sigma_{ij}\sigma_{kl} - h\sum_{ij}\sigma_{ij},
\end{equation}
where $J=-1$ is the antiferromagnet coupling constant and $h$ is the external magnetic field. The first summation is restricted to nearest neighbor spins, and the second summation is over all spins in the system. The dynamic is chosen to be the single spin flip Glauber dynamic \cite{Glauber1963time}. The rate of flipping a spin is given by 
\begin{equation}
    R_{\textrm{flip}}(T_b) = \frac{1}{1+e^{\Delta E / T_b}},
\end{equation}
where $\Delta E$ is the energy increment due to the flip.

As already mentioned, there is no hope to find ${\bf v}_2$ numerically, even for a moderate case of $N=70$, corresponding to $2^{4900}$ microstates. To project ${\bf p}(t)$, we thus choose two observables: the mean and staggered magnetization, defined for a microstate by
\begin{eqnarray}
    M = N^{-2}\sum_{ij}\sigma_{ij};\;\; S = N^{-2} \Biggl|{\sum_{ij}s(ij)\sigma_{ij}}\Biggl|,
\end{eqnarray}
where $s(ij) = 1$ ($-1$) for even (odd) value of $i+j$, specifying two sub-lattices, and the absolute value in $S$ is used since the two sub-lattices are symmetric due to periodic boundary conditions.

\begin{figure}
    \centering
    \includegraphics[width=1\columnwidth]{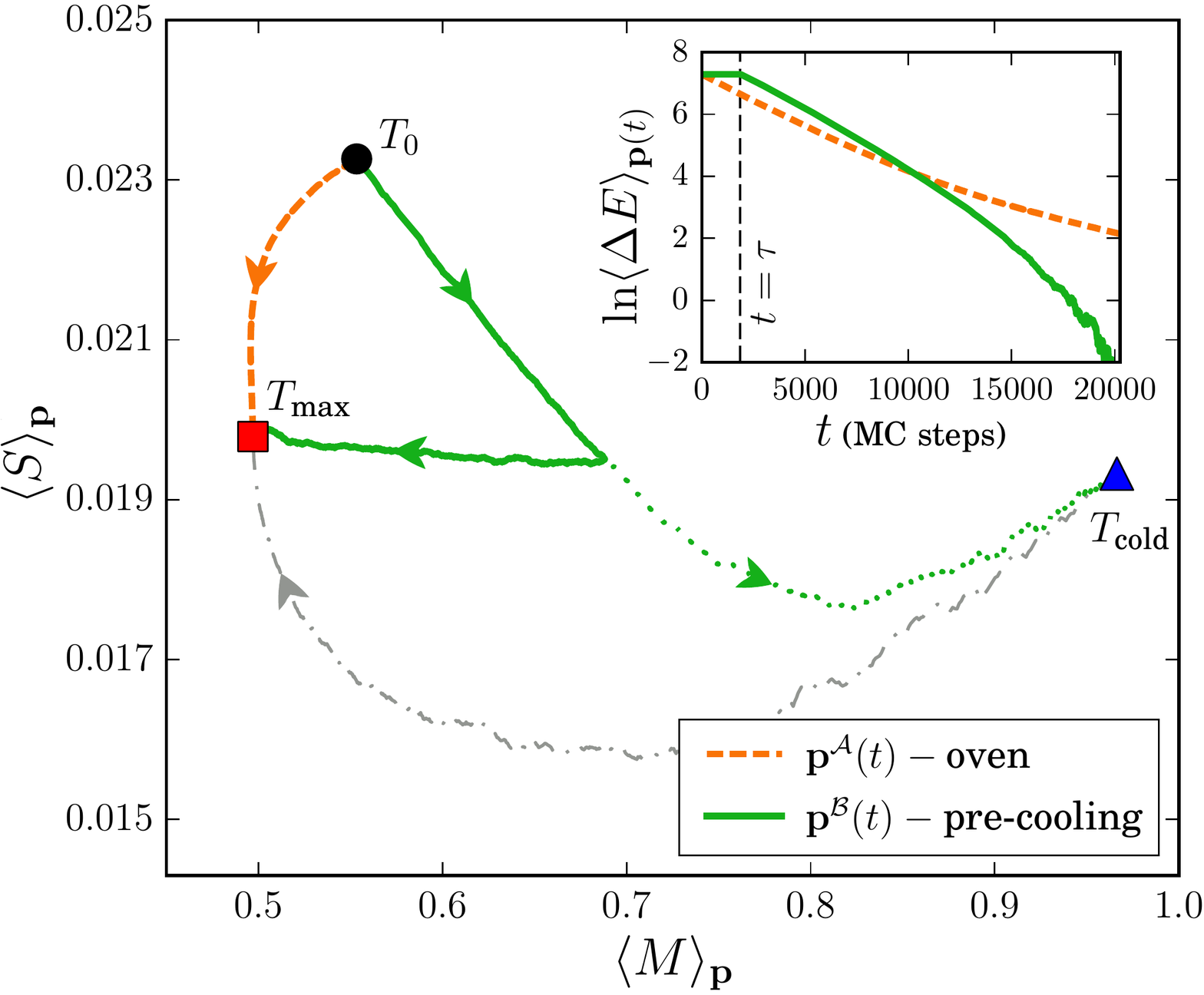}
    \caption{
    An example for a pre-cooling protocol in the 2D Ising model with external magnetic field $h=1.0025$. The black point, red square and blue triangle are the mapped equilibrium points $(\langle M\rangle_{\boldsymbol{\pi}(T)},\langle S\rangle_{\boldsymbol{\pi}(T)})$ at the initial temperature $T_0=0.125$, the maximal $T_\textrm{max}=0.625$ and the cold $T_{\textrm{cold}}=0.0125$, respectively. 
    The orange dashed line is the trajectory of the oven protocol, initiated at temperature $T_0$ and approaches the $T_\textrm{max}$ equilibrium from the projection of ${\bf v}_2$ direction. Another trajectory (gray dot-dashed line), is initiated at $T_{\textrm{cold}}$, evolves under the oven protocol and approaches the same equilibrium from an opposite direction.
    The green solid line is the trajectory of the pre-cooling protocol, with a pre-cooling duration of $\tau=1850$ MC steps, which approaches the $T_\textrm{max}$ equilibrium from a different direction -- and thus from the fast manifold. The green dotted line is the trajectory the system would have followed had it stayed coupled to $T_\textrm{cold}$.
    Inset: the energy difference between the energy ${\bf p}$-averaging of a trajectory and equilibrium, plotted for the oven and pre-cooling protocols.}
    \label{fig:Ising}
\end{figure}

Finding a pre-cooling protocol in the 2D Ising model described above with $N=70$ is demonstrated in Fig.~(\ref{fig:Ising}). The projected trajectories $(\langle M\rangle_{{\bf p}(t)},\langle S\rangle_{{\bf p}(t)})$ were calculated using $10^7$ realizations of a MC simulation. To sample the realizations from $\boldsymbol{\pi}(T_0)$, the state of every spin in each realization was chosen randomly, and the Glauber dynamic was applied to all the realizations for $10^6$ MC steps, with $T_b=T_0$. After preparation, the oven protocol, denoted by $\A$, was applied and $(\langle M\rangle_{{\bf p}^\A(t)},\langle S\rangle_{{\bf p}^\A(t)})$ was measured. 
As Fig.~(\ref{fig:Ising}) shows, another trajectory (gray dot-dashed line), where the system is prepared at $T_b=T_\textrm{cold}$ and evolves under the same protocol, approaches the same equilibrium point from an opposite direction. This implies that the sign of $a_2$ in the cold initial condition is opposite to that of the hot temperature.
The pre-cooling protocol, denoted by $\B$, was found by choosing $\tau$ such that the corresponding trajectory $(\langle M\rangle_{{\bf p}^\B(t)},\langle S\rangle_{{\bf p}^\B(t)})$ approaches the equilibrium point from a different direction. This different direction cannot be the projection of ${\bf v}_2$, and therefore the corresponding trajectory must be on the fast manifold.



To demonstrate that the pre-cooling protocol is indeed faster, we used the ${\bf p}$-averaged energy difference to measure the distance of a trajectory to equilibrium, as suggested in \cite{SpinGlassMpemba}. This distance function captures the faster relaxation rate of the pre-cooling protocol, compared to the oven protocol, as shown in the inset of Fig.~(\ref{fig:Ising}). In the Supplemental Material \cite{SI} we show that the same behaviour appears in the thermodynamic limit of the mean-field antiferromagnet Ising model too.

In this manuscript we have demonstrated how heating processes in several types of systems can be exponentially improved using a pre-cooling strategy. Optimal cooling protocol can similarly be found using the same approach. Our method is based on the cooling and heating dynamics in probability space, which can be immense, but as we demonstrated in the 2D Ising system, projecting it to a lower dimension space enables to apply it even in many-body systems. The ability to choose the observables on which the dynamic is projected suggests that this method can be used in experiments too.

\begin{acknowledgments}
    O.R. is the incumbent of the Shlomo and Michla Tomarin career development chair, and is supported by a research grant from Mr and Mrs Dan Kane and the Abramson Family Center for Young Scientists. We would like to thank H. Aharoni, Prasad V. V., O. Hirschberg, M. Vucelja and I. Klich for useful discussions.
\end{acknowledgments}

%

\end{document}